\documentstyle[12pt,aasms4,flushrt]{article}

\tighten

\input psfig

\lefthead{Allende Prieto, Garc\'{\i}a L\'opez, Lambert \& Gustafsson}
\righthead{Spectroscopic observations of  convective patterns}
\begin{document}

\title{Spectroscopic observations of  convective patterns in the atmospheres of metal-poor stars}

\author{Carlos Allende Prieto\footnote{Present address: McDonald Observatory and Astronomy Department,  
The University of Texas at Austin, RLM 15.308, Austin,
TX 78712-1083,  USA} and Ram\'on J. Garc\'{\i}a L\'opez}
\affil{Instituto de Astrof\'\i sica de Canarias \\ E-38200, La Laguna,
Tenerife,  \\  SPAIN}

\author{David L. Lambert} \affil{McDonald Observatory and Department of
Astronomy, \\ The University of Texas at Austin \\ RLM 15.308, Austin,
TX 78712-1083, \\ USA}

\author{Bengt Gustafsson}
\affil{Uppsala Astronomical Observatory \\ Box 515, S-751 20 Uppsala \\ SWEDEN}

\authoraddr{IAC, E-38200, La Laguna, Tenerife,  SPAIN}

\authoremail{callende@iac.es, rgl@iac.es, dll@astro.as.utexas.edu and bg@astro.uu.se}  

\begin{abstract}

Convective line asymmetries  in the optical spectrum of two metal-poor
stars, Gmb1830 and HD140283, are compared to those observed for solar
metallicity stars.  The line bisectors  of the most metal-poor star,
the subgiant HD140283, show a significantly larger velocity span that
the expectations for a solar-metallicity star of the same spectral type
and luminosity class. The enhanced line asymmetries   are interpreted
as the signature of the  lower metal content, and therefore opacity, in
the convective photospheric patterns. These findings point out the importance of the three-dimensional convective velocity fields
 in the interpretation of the observed line asymmetries in metal-poor stars, 
and in particular, urge for caution when deriving isotopic ratios 
from  observed line shapes and shifts using one-dimensional 
model atmospheres.

The mean line bisector of the  photospheric atomic lines is compared
with those measured for  the strong  Mg I b$_1$ and b$_2$ features. The
upper part of the bisectors are similar, and assuming  they overlap,
 the bottom end of the stronger lines, 
which are formed higher in the atmosphere, goes
much further to the red.  This is in agreement with the expected
decreasing of the convective blue-shifts in  upper atmospheric layers,
and compatible with the high velocity redshifts observed in the
chromosphere, transition region, and corona of late-type stars.

\end{abstract}
 

\section{Introduction} 
\label{sec1}

The solar granulation pattern observed by direct imaging in the optical
continuum is the result of the convective motions in the solar
envelope. The velocity fields and spatial patterns present in the solar
photosphere leave a signature on spectra, even when they lack spatial
resolution. Line asymmetries  reveal similar shapes for most  lines
(Dravins, Lindegren \& Norlund 1981), while line shifts
 become gradually bluer when the line formation occurs deeper in the
photosphere (Allende Prieto \& Garc\'{\i}a L\'opez 1998a, hereafter
APGL).  These features cannot be realistically explained by any other
known mechanism such as isotopic shifts, hyperfine structure, pressure
shifts, or line blends.

Similar effects are expected to be present
 in  other stars. Over the last two decades, David Gray and
collaborators (see, e.g., Gray 1982, Gray \& Toner 1986, Toner \& Gray
1988, Gray \& Nagel 1989, Gray et al. 1992) and Dainis Dravins (see,
e.g., Dravins 1982, 1987a, 1987b) have extended the measurement of line
asymmetries  to many other stars, confirming
 the expectations:  the shapes of the line bisectors of late-type stars
with convective envelopes are similar to the solar
 case.  Surprisingly, opposite curvature are found in the bisectors of
hotter stars, which are not expected to develop convective envelopes.
The  difficulties in obtaining highly accurate radial velocity
measurements and the need to  separate
 the radial velocity, the gravitational
 shift, and convective shifts has precluded the use of
  absolute line shifts as a tool to probe convection on late-type stars.
 The use of differential line shifts has already been attempted by
 Nadeau \& Maillard (1988)  for M giants.

Gray (1982) and Gray \&  Toner (1986) identified a sequence in the line
asymmetries for late-type dwarfs, giants, and supergiants. Using no
information about the line shifts, they averaged line bisectors for lines of
different depths at the line center,  showing that this method is very
useful as a first order approach to understanding the changes of the
granular patterns with atmospheric parameters.

In the last two decades, large advance has been made in the 
modeling of convection
in stellar atmospheres. Three-dimensional
 hydrodynamical numerical simulations of the
stellar atmospheres are able now to reproduce the observed line
asymmetries, opening up the possibility of understanding convective
velocity
 fields in the photospheres of stars others than the Sun (see, e.g., Nordlund \&
 Dravins 1990).  Although empirical models  of  granulation can be
 constructed and may be very useful to disentangle the interplay
between velocity fields and granulation contrast,  full
 hydrodynamical modeling is the most powerful tool for understanding
the  physical mechanism behind the observed
convective patterns.
 
Of particular significance is the proper understanding of convection
and the effects of convective inhomogeneities in metal-poor stars on
the main sequence or close to it (subdwarfs, subgiants). Abundance
analyses of such stars are fundamental in the study of primordial
nucleosynthesis as well as the early evolution of the Galaxy. In these
stars, the convection zones and the resulting inhomogeneities reach
visible photospheric layers, mainly due to the high transparency of the
gas because of the low electron pressure and the lack of metal
absorption in the ultraviolet. The latter circumstance also leads to a
hot non-local radiation field in the near ultraviolet which may induce
severe departures from local thermodynamic equilibrium (LTE). It is 
thus very important to explore the effects of convection and departures 
from LTE in stars of this type.

Detailed comprehension of surface convection 
in metal-poor stars is of high importance for fine
analysis of spectral line shapes, such as  the retrieval of 
isotopic ratios of lithium (Smith, Lambert \& Nissen 1998), 
boron (Rebull et al. 1998), or barium (Magain \& Zhao 1993), which
have to rely on one-dimensional models that may introduce an important
uncertainty.

 We have  acquired  spectra  of   adequate quality 
of  two metal-poor stars,  to address   these fundamental 
questions of whether,  or not,  the three-dimensional inhomogeneities and
 the convective velocity motions in the photospheres   of metal-poor
stars lead to severe errors in the results obtained from the   use of
one-dimensional model atmospheres in the spectroscopic analysis of  of
metal-poor stars.   In this paper,  we
 select clean profiles,  measure,  and average line bisectors of
 different lines in order to compare  convection in the photospheres
 of  solar composition and  metal-poor stars. After giving the details
 of the observing and reduction procedure in \S2, we shall discuss the
  solar case and carefully check the quality of the spectroscopic
  observations
   in \S3. The  analysis of line asymmetries in the photospheres of the
   metal-poor  compared to the solar composition stars is the subject
   of \S4, and \S5.

\section{Observations}
\label{sec2}

We have selected two well-known field stars belonging to the 
population II: the moderate metallicity dwarf Gmb1830
(HD103095, HR4550; [Fe/H] $\sim$ --1.3; G8 V) and the 
more metal-poor subgiant HD140283 ([Fe/H] $\sim$ --2.7; G0 IV). 
The Sun and the solar-like metallicity star $\theta$ UMa (HD82328, HR3775;
F6 IV)  were included in the program to be used as  references.

Observations were carried out during three campaigns from 1995 to 1997
using the {\it 2dcoud\'e} echelle spectrograph (Tull et al. 1995)
coupled to the Harlan J. Smith 2.7m Telescope at McDonald Observatory
(Mt. Locke, Texas). The cross-disperser and the availability of a
2048x2048 pixels CCD detector made it possible to gather up to 300
\AA\ in a single exposure, in a series of non-overlapping segments. The
set-up provided resolving powers ($\lambda/\Delta\lambda$) in the range
170,000-220,000.  As many 1/2 hour exposures were acquired as were
needed to reach a final signal to noise ratio (SNR) of $\sim$ 300--600.
Table 1 describes the three observational campaigns  devoted to the program.

A very careful data reduction was applied using the 
IRAF\footnote{IRAF is distributed
 by the National Optical Astronomy Observatories, which are 
operated by the Association of Universities for Research in 
Astronomy, Inc., under cooperative agreement with the 
National Science Foundation.} software
package, and consisted in: overscan (bias) and scattered light
substraction, flatfielding, extraction of one-dimensional spectra,
wavelength calibration, and continuum normalization. Wavelength
calibration was performed for each individual image on the basis of
$\sim$ 300 Th-Ar lines spread over the detector. The possibility of
acquiring daylight spectra with the same spectrograph 
allows us to perform a few interesting
tests. Comparison of the wavelengths of 60 lines in a single daylight
spectrum (SNR $\sim$ 400-600, depending on the spectral order)  with the
highly accurate wavelengths measured in the solar flux spectrum by
Allende Prieto \& Garc\'{\i}a L\'opez (1998b) showed that the rms
differences were at the level of 58 m s$^{-1}$ ($\sim \frac{1}{11}$ pixel).

Before coadding the individual one-dimensional spectra, they were first
cross-correlated to correct for the change in Doppler shifts and
instrumental displacements, such as those produced by the variation of
weight as the CCD's liquid nitrogen dewar empties. Fig. 1  shows
the measured shifts between different spectra of HD140283, relative to
the first of them, on the night of May 20 1995. The observed
shifts (joined by the solid line) do not correspond to those expected 
from the difference of velocities in the line of  sight between 
the Earth and the Sun,
indicated by circles in Fig. 1.  Our procedure introduced an 
uncertainty in the wavelength scale, whose magnitude depends on the
SNR, the presence of telluric lines, and the time separation among the
individual spectra. When the standard deviation of the velocity shifts
from the cross-correlation of the different available orders was below
$\sim$ 150 m s$^{-1}$, the frames were co-added to increase the SNR.
This strategy ensures that the errors introduced in the line shifts
when co-adding the
spectra are of the order of $\frac{150}{\sqrt{N}}$ m
s$^{-1}$ or less, where N, the number of useful orders, 
is in the range 8--17. In this way no significant extra asymmetry is 
artificially produced.

Fig. 2 demonstrates how well the shifts are determined for the individual
spectra of HD140283 taken on May 20 1995. The left panel shows the 
Fe I line at 5393 \AA\ in the different spectra, after correcting the shifts 
displayed in Fig. 1, while the right panel shows the resulting pattern, ten
times magnified, after  subtracting the mean spectrum. It is apparent in this
 figure that  no significant residuals remain.

A final test to   check that our procedure leads to
consistent results is to compare   the bisectors measured in the
individual exposures with the averaged bisector, and with the bisector
measured in the averaged spectrum. This is shown in Fig. 3, for the 
series in Fig. 1. The average bisector (thick solid line)
is in agreement with the bisector of the average line profile (dashed
line). The bisectors were  averaged using normalized weights $\propto
\frac{1}{\sigma^2}$, which corresponds to the weighting performed when
co-adding the spectra. While the direct average of  the bisectors
relative to the line center avoids the shifting of the
spectra to the same zero, it requires the location of the centers of the line
profiles in the noisy individual exposures, increasing the errors in the final
bisector. Alternatively, the final bisector can be measured directly on the 
averaged spectrum, computed after shifting safely the individual spectra to
a common zero through  simultaneous cross-correlation of all the
spectral orders recorded in a given image. The latter procedure has been 
adopted in this work.

\section{The Sun: accuracy of the measurements}

The highest quality  high resolution stellar spectra currently
available are those of the Sun. The study of line asymmetries and
shifts in the optical solar spectrum has been the subject of much work
since the early studies of St. John (1928) and Burns and collaborators
at Allegheny Observatory (Burns 1929, Burns \& Kiess 1929, Burns \&
Megger 1929).  There are recent measurements 
in the center of the solar disk, at
 different positions across it, in total flux, and with high spatial
and time resolution (see, e.g., Pierce \& Breckenridge  1973,
  Dravins, Lindegren \& Norlund 1981, Livingston 1982, Balthasar 1984, 
Brandt \& Solanki 1990,  Stathopoulou \&
Alissandrakis 1993, APGL), but
 a comprehensive study and classification of the line asymmetries in
the flux spectrum of the Sun is still missing. Such a study providing
 an average solar bisector  could serve as a template for comparison
 with  other stars.

Making use of the solar flux spectrum in the atlas of Kurucz et al.
(1984), we have measured the line asymmetries of 39 Fe I clean lines selected 
by Meylan et al. (1993) from the fit of Voigt profiles in the same
 atlas. The line asymmetries were quantified by means of the bisectors
 at a given set of flux levels from the continuum. 
 Irregular parts  in the line profiles were removed and excluded 
from the average.  All the lines were 
averaged together, to produce a mean flux bisector, 
as commonly done with stellar bisectors (Fig. 4; solid line with error bars
indicating the mean error).  A new average bisector  (Fig. 4; solid line 
with shadow  area indicating the mean error) was constructed  taking 
into account the line shifts for Fe I lines
measured by APGL from the same atlas, to place the individual bisectors
on an absolute scale. In Fig. 4, the {\it absolute} mean flux 
bisector has been shifted from its true (bluer) position to overlap 
with the mean flux bisector computed
without taking into account the blue-shifts of the individual lines. 
The differences between the profiles with and
without correction for the velocity shifts are comparable to the
precision with which bisectors are measured, but more remarkable, the
velocity difference or span between the bluest and the 
reddest part of the mean bisectors does not change.
 
 As listed in Table 1, we have acquired day-light spectra
  in different spectral ranges. Figure 5 compares the  bisector
 measured in these spectra (curve with error bars) to that measured  in the
 solar FTS atlas of Kurucz et al. (1984) from three  Fe I lines
 ($\lambda\lambda$ 5217, 5296, and 5933 \AA). The agreement
  is very good, indicating, as expected, that the errors in our
 measurements are below $\sim$ 60 m s$^{-1}$. This comparison
warranties the adequateness of the procedure employed to remove
 the scattered light, as this problem is much more important for
the day-light spectra than for the stellar spectra. The average of 14 lines
 in the McDonald day-light spectrum provides a {\it solar flux mean
 bisector} quite similar to that previously calculated from 
 the FTS atlas (see below). Here and hereafter, error bars for the bisectors
of individual lines are computed following the considerations in 
Gray (1983).

 The convective line asymmetries depend on the velocity fields, the
 granulation contrast,  and the area occupied by the rising granules
 and the falling intergranular
  lanes in the line formation layers. As a result of this, convective
  line asymmetries have been observed to vary with the  line depth, the
 atomic parameters (chemical species, excitation potential, transition
 probability, etc.), i. e.,  all those parameters defining the height
 of the line formation region in the atmosphere. Some spectral lines
 can be affected by other sources of asymmetry, such as isotopic shifts
 or hyperfine structure splitting. These are difficult to estimate as
 accurate laboratory data are missing (see, e.g. Kurucz 1993). Lacking
laboratory studies and accurate calculations on collisional broadening,
we can not definitely exclude that some lines could be affected by
pressure shifts, inducing line asymmetries  (Allende Prieto,
Garc\'{\i}a L\'opez \& Trujillo Bueno 1997).  These facts  
 make the use of  average stellar bisectors questionable. Ideally, detailed
 modeling of the convective line asymmetries should be carried out by
  comparison of predicted and observed  profiles of individual lines.
However,  there is a common behavior of the line
asymmetries
 for most of the  lines in the solar and stellar spectra and, at least
for the solar case, the velocity span of the mean bisector does not
change much between taking into account or not the absolute position
of the individual lines. Furthermore,  there is a remarkably
smooth variation of the average
 stellar bisectors with spectral type and luminosity class (Gray 1982,
 Gray \&  Toner 1986). This  justifies the use of mean
  bisectors as a first approach to the classification of the line
 asymmetries  across the HR diagram.

\section{Metal-poor stars. Photospheric line asymmetries.}

For the most metal-poor star in the sample (HD140283) the task of
selecting lines free of blends and anomalous shapes is straightforward,
because the lack of metals greatly reduces the overlapping of different
lines. For the less metal-poor stars,  blending becomes more common,
and the selection more difficult.   The calculation of an initial 
average and its standard deviation was made
and then a new average, excluding the points deviating significantly
from the first average, was calculated.  Finally, it was verified that
a more critical selection, keeping only the bisectors which exhibited
the dominant shape, in all cases, provided almost identical results.

\subsection{The dwarf Gmb1830}

We have identified a total of 25 clean lines in the  spectral range
available for Gmb1830 (G8V). Their wavelength, suggested identification,
equivalent width, and excitation potential are listed in Table 2.  We
proceeded as described in the preceding section for the Fe I lines in
the solar atlas, and obtained the mean flux bisector for Gmb1830 and
the sky-light spectra acquired at McDonald. They are listed in Table 3.
Figure 6a shows all the bisectors measured in Gmb1830, and the mean
flux bisector for this star, surrounded by error bars, describing the
mean error. Figs. 6b is similar to 6a, but for
 the Sun (from 14 clean lines observed at McDonald). The line bisector
shapes are not highly homogeneous  but the typical C shape, which can
be attributed to the effects of  convection, is apparent for most of
the lines.

Line bisectors for F-K dwarfs behave in such a way that, although the
solar-like C-shape is common for all, the velocity span shows the
smallest values around the spectral class G8 (Gray 1982). Direct
 comparison of the mean bisectors in Figure 7 shows  that the mean
bisector of Gmb1830 (mean errors represented with error bars) 
shows a smaller velocity span than that of a {\it solar metallicity} 
G2 (the Sun; mean errors in gray). This implies that, for this 
moderately metal-poor dwarf, we  do not detect any significant  
signature of the lower metallicity  in 
its mean flux bisector.

Rotation strongly affects the shape of the integrated-light line
asymmetries (see Gray 1986; Smith, Livingston \& Huang 1987; Dravins \&
Nordlund 1990).  However, the rotational velocities of the Sun (1.9 km
s$^{-1}$; Gray 1992) and Gmb1830 (2.2 km s$^{-1}$; Fekel 1997) are
quite close, and  then, their asymmetries can be directly compared.
Several other factors may be playing a significant role in this
comparison.  The presence of one or more unresolved companions could
well induce systematic line asymmetries in the spectral lines that
might be wrongly interpreted as convective patterns. Beardsley,
Gatewood, \& Kamper (1974) claimed a detection of radial velocity
variations in the spectra of Gmb1830, but this claim was not confirmed
(Griffin 1984; Heintz 1984). The {\it Hipparcos} catalogue (ESA 1997)
has found the star to show photometric variations spanning 0.14 mag.,
and although the same catalogue does  not register a visual companion
 for Gmb1830 within 10 arcseconds, it has been claimed in the past (see
Beardsley et al. 1974) that the star
 has a fainter (5-5.5 mag.) companion.

Cyclical activity and magnetic  variations are known to be linked to changes 
in oscillations, and granulation properties (Gray \& Baliunas 1995; 
Jim\'enez-Reyes et al. 1998).  Radick et al. (1998) have detected 
solar-like periodic variability in the Ca II H and K emission of Gmb1830 with
a period close to seven years. The possible differences between the 
mean bisector of Gmb1830 and the Sun associated with their different 
metallicities, may indeed have been  diluted under some of these effects.
 
\subsection{The subgiant HD140283}

As a result of the  very low metal content of the star, the high resolution
spectrum of HD140283 shows only a few lines. The practical advantage is
that the lines present are quite clean. A total of 24 lines were
selected in the three spectral ranges available for HD104283 (G0 IV),
while only 16 were considered clean in the case of $\theta$ UMa (F6 IV), 
a comparison {\it solar metallicity}  star. The data on the lines selected
is also included in Table 2. 
All the  measured bisectors in these
stars are plotted in Figure 8,  the
 mean flux bisector is overplotted with the mean error marked by
the error bars. Table 4 lists the flux mean bisectors.  Almost every
line detected in the spectrum of HD140283 was considered clean.
The homogeneity of the bisector shape is higher than for the  cooler dwarfs
studied in the preceding sub-section.

Fig. 9 directly compares  the flux mean bisectors for the two stars and
the Sun (solid line with the mean error marked in grey). The
expectation based on the smooth trends with the
 spectral class found by Gray (1982) is that the hotter star ($\theta$
 UMa; F6 IV; dashed line with error bars) should have  photospheric
bisectors with the larger velocity span. That is in clear
 contradiction with Fig. 9, where the bisector corresponding to
 HD140283 (solid line with error bars) shows a red asymmetry as high as
$\sim$ 300 m s$^{-1}$. The obvious suggestion is that the
 abnormal behavior of the line bisectors measured in the spectrum of
HD140283  is the result of its very low
 metallicity (a factor $\sim$ 500  less  than the Sun or $\theta$ UMa).

In this case there is a significant difference in the rotational
velocities between HD140283 ($\le$ 3.5 km s$^{-1}$; Magain \& Zhao
1993) and $\theta$ UMa (6.4 km s$^{-1}$; Fekel 1987). This difference 
 is  large enough to produce a
respectable effect.  However, the comparison with the slower Sun (G2
V\footnote{Differences between luminosity class V and IV are likely to
be negligible (see  Gray 1982).}) 
 is  not affected by this parameter, and leads to the same result.

The time span of our observations (three years) suggests that the line
asymmetries observed in HD140283 are stable. The star has been
monitored for radial velocity variations with a negative result (Carney
\& Latham 1987, Mazeh et al. 1996). $\theta$ UMa has been claimed to
show periodic radial velocity variations by Abt \& Levy (1976), but this
has been placed into question by the analysis of Morbey \& Griffin (1987).

\section{Mg I b$_1$ and b$_2$ lines in the spectrum of HD140283}

Mean flux bisectors studied in \S 4  correspond to photospheric lines,
and characterize the velocity fields only in this region of the atmosphere. 
However, velocity fields in upper layers, such as the 
chromosphere (Samain 1991, Garc\'{\i}a L\'opez et al. 1992), 
the transition region, or the corona (Brekke et al. 1998) 
of late-type stars have been shown 
to be much stronger. While the photospheric
observed line shifts are directed bluewards and amount 
a few hundreds meters per  second, transition region and
 coronal lines are shifted 
to the red by kilometers per  second (Wood et al. 1996, 
Wood, Lynsky \& Ayres 1997).

We have measured the asymmetries of the strong MgI b$_1$ and b$_2$
lines at 5183 and 5172 \AA, respectively, whose cores are known to
 form higher up than the photospheric layers in the solar atmosphere,
in the spectrum of HD140283.  The line bisectors are  displayed in
Fig.10, and compared with the mean flux bisector. In this Figure, the
zero velocity is again arbitrarily set to the bottom of the lines. They
exhibit a  C-shape, although are quite different to the bisectors of  the
 photospheric lines. This could indicate a larger dissimilitude,
compared with the photospheric lines,
 between the velocity fields and the inhomogeneities in the layers
where the core and the wings are formed. Assuming  the upper part of
the photospheric and the strong-line bisectors overlap (as suggested by
the shape and the velocity span),  the excursion of the line bottom to
the red would be tracing the disappearance of the photospheric
correlation between  temperature and velocity, and therefore the convective
blueshift,  towards higher atmospheric layers.

Alternatively, the peculiar shape of these bisectors
might be the result of the presence of significant contributions of
different isotopes of magnesium.  There are three magnesium isotopes in
the solar-system mixture, $^{24}$Mg, $^{25}$Mg, and $^{26}${Mg}, whose
abundance fractions are 0.7899, 0.1000,
 and 0.1101 (Anders \& Grevesse 1989), but it is established that the
 fractions of  $^{25}$Mg and $^{26}$Mg, relative to $^{24}$Mg decline
with metallicity (Barbuy, Spite \& Spite 1987, McWilliam \& Lambert
1988), and the resulting asymmetry
 is expected to be very small for a star like HD140283.

Unlike  the solar metallicity stars,  the lack of line crowding makes
it possible to measure the asymmetries of lines which form between
photospheric and chromospheric layers in the extreme metal-poor stars.
This could be an important tool to understand how the  dynamics of the
atmosphere changes from producing blueshifts in the photosphere to
redshifts in higher layers. The observations obviously must constrain
future three-dimensional simulations of photospheric dynamics further
out from the center of the star.

\section{Summary and conclusions}

We have searched for differences between the convective velocity fields and
granular motions  of metal-poor stars and solar metallicity stars by
observing line asymmetries   in the optical spectra
at very high resolution.

Clear differences have been found for the most metal-poor star in our
sample, probably reflecting the low 
opacity of the metal deficient atmospheres and the changes in visible 
convective flow patterns due to this.  The line asymmetries found in this case
show a significantly different
shape as compared with its solar-metallicity counterpart, 
perfectly distinguishable 
from the observed line-to-line differences.

The lack of metal line blends and the relatively narrow line wings in
metal-poor stars
 makes it possible to measure the line asymmetries in strong lines such
 as the Mg I b$_1$ and b$_2$, whose cores are formed higher in the
atmosphere, possibly revealing a  convective pattern rapidly  changing
with depth which shows up as a markedly redder asymmetry in the line
core, as compared with photospheric lines.

Detailed comparison between observed line profiles and
three-dimensional numerical simulations of the photospheres of
late-type stars, as affected by the underlying convective dynamics,
should be a  powerful tool to improve the understanding of the
atmospheric structure and dynamics of these objects (Allende
Prieto et al. 1999, Asplund et al. 1999). Such comparisons should
 give place to more reliable abundance analyses for these stars.

We thank the staff at McDonald Observatory, in particular David Doss,
 for their kind and professional help.  The comments of the referee
were particularly helpful to improve some aspects of both the contents
and the presentation. This work has been partially funded by the
Spanish DGES under projects PB92-0434-C02-01 and PB95-1132-C02-01, the National Science Foundation (grant AST961814),  and the Robert A. Welch
Foundation of Houston, Texas.

\clearpage

\centerline{\bf FIGURE CAPTIONS}

\bigskip

\figcaption{Relative displacements in the spectral direction
 between different exposures of HD140283 on May 20, 1995 (crosses joined
by the solid line), and expected shifts due to the Earth-Sun motion (circles).}

\figcaption{Left panel: individual spectra of the Fe I $\lambda$5393 \AA\ 
line in the spectrum of the star HD140283 on May 20
 1995, after removing the relative shifts. The radial velocity has not
been corrected. Right panel: remaining residuals after subtracting the mean
profile to the individual spectra in the left panel (the vertical scale has been
magnified by a factor of ten).}

\figcaption{The average of the bisectors measured in the individual 
spectra  of HD140283 shown in Fig. 2 (thick solid line), whose
shifts are tracked in Fig. 1, 
is in perfect  agreement with the bisector  measured in the
 average profile, constructed 
after correcting the relative shifts (dashed curve). The 
thin solid lines show the individual bisectors.}

\figcaption[]{The  average solar line bisector, constructed from 39
{\it clean} Fe I lines, with the individual line shifts taken into
account (mean errors in grey) and with line shifts not considered (line
with error bars).}

\figcaption[]{Comparison between the solar bisectors 
(Fe I, $\lambda\lambda$ 5217, 5296, and 5933 \AA), as measured
in the FTS atlas of Kurucz et al. (1984; $\sigma$ represented
by the shadow), with those
 from the day-light spectra acquired at McDonald Observatory 
( $\sigma$ represented by error bars).}

\figcaption[]{All the measured bisectors  for individual atomic lines
in the spectra of the  Gmb1830 (a), and  the Sun (b) are plotted (thin lines). The thick lines represent the {\it mean flux bisectors}, and the error bars
indicate the mean error at a given normalized flux.}

\figcaption[]{The mean flux bisector of the metal-poor dwarf Gmb1830
(G8; solid line with error bars)  as compared with the
hotter Sun (G2; solid line with the mean errors in gray).}

\figcaption[]{All the measured bisectors in the spectra of HD140283 (a),
and  $\theta$ UMa (b) (thin lines). The
thick lines represent the {\it mean flux bisectors}, and the error bars
indicate the mean error at a given normalized flux.}

\figcaption[]{The line bisectors' shape in the spectrum of HD140283 (G0;
solid line with error bars). The {\it mean flux bisector}
for the {\it solar-metallicity} comparison star $\theta$ UMa (F6; dashed line) and the Sun (solid line with mean errors in grey) 
are also shown. The bisector of HD140283 shows a larger velocity 
amplitude than the metal-rich stars do.}

\figcaption[]{The bisectors of the strong Mg I b$_1$ (deep bisector with error
bars) and b$_2$ (deep bisector with $\sigma$ in shadow) lines
(solid lines), are compared with the flux mean  bisector of HD140283. 
The core of the Mg b lines  are formed
higher in the atmosphere than the photospheric lines employed for
constructing the flux mean bisector as suggested by the remarkable
asymmetry to the red, which reveals a change in the velocity-temperature
pattern, such as that observed in the solar chromosphere. The absolute position of the bisectors is arbitrary, as the line bottom is assumed to be
 at zero velocity.}

\clearpage

\begin{deluxetable}{cccc}
\tablecaption{Observations: Dates, Spectral Ranges and Signal-to-noise Ratios. \label{table1}}
\tablehead{\colhead{Star} & \colhead{Spectral  range (\AA)\tablenotemark{a}} & \colhead{Date} 
& \colhead{SNR} }
\startdata 
Sun	&	4853-6404  &	20-May-95 &	$>$ 500 \\
 "	&	4853-6288  &	21-May-95  &	" \\
"	&	5204-7037 &	27-Feb-96 &	" \\
"	&	5549-7875  &	23-May-97  &	" \\
"	&	4372-5434 &	24-May-97 &	" \\
"	&	5687-8156 &	24-May-97 &	" \\	
 "	&        5687-8156 &	25-May-97 &	"	\\
Gmb1830 &	4853-6404 &	26-Feb-96 &	250-350 \\
"	 &	5204-7183 &	27/29-Feb-96 &	300-400 \\
HD140283 &	4850-6400  &	19-May-95 &	150-250	 \\
"	 &	4850-6400 &	20-May-95 &	300-500 \\
"	 &	4850-6400 &	21-May-95 &	200-300 \\
"	 &	4424-5497  &	26-Feb-96 &	70-100 \\
"	 &	4372-5434 &	23/24/25-May-97 & 500-650 \\
$\theta$ UMa	 &	5690-8360 &	26-Feb-96 &	600 \\
\enddata
\tablenotetext{a}{The specified spectral range is not continuous, 
as the different spectral orders do not overlap.}
\end{deluxetable}


\begin{deluxetable}{rrrr}
\tablecaption{Atomic Spectral Lines Measured \label{table2}}
\tablehead{\colhead{Wavelength (\AA)}    &  \colhead{Species} &  
 \colhead{W$_{\lambda}$ (m\AA)}   & 
\colhead{E.P. (eV)} }
\startdata
\cutinhead{ Sun (skylight)}
4950.11  & Fe I   &   96  &  3.42   \nl
5001.87  & Fe I   &   343  &  3.88   \nl
5217.40  & Fe I   &   120  &  3.21   \nl
5296.70  & Cr I   &   98  &  0.98   \nl
5415.21  & Fe I   &   191 &  4.39   \nl
5576.10  & Fe I   &   153  &  3.43   \nl
5701.56  & Fe I   &   84  &  2.56   \nl
5852.23  & Fe I   &   43  &  4.55   \nl
5933.80  & Fe I   &   105  &  4.64   \nl
6151.62  & Fe I   &   50  &  2.18   \nl
6166.44  & Ca I   &   77  &  2.52   \nl
6335.34  & Fe I   &   100  &  2.20   \nl
6336.83  & Fe I   &   126  &  3.69   \nl
6498.95  & Fe I   &   138  &  0.96   \nl
\cutinhead{ Gmb1830}
4999.51  & Ti I   &   162  &  0.83   \nl
5001.87  & Fe I   &   219  &  3.88   \nl
5083.34  & Fe I   &   130  &  0.96   \nl
5215.19  & Fe I   &   98  &  3.26   \nl
5217.40  & Fe I   &   97  &  3.21   \nl
5225.53  & Fe I   &   59  &  0.11   \nl
5229.86  & Fe I   &   104  &  3.28   \nl
5232.95  & Fe I   &   347  &  2.94   \nl
5302.31  & Fe I   &   125  &  3.28   \nl
5307.37  & Fe I   &   65  &  1.61   \nl
5379.58  & Fe I   &   22  &  3.69   \nl
5381.03  & Ti II   &   26  &  1.57   \nl
5389.49  & Fe I   &   52  &  4.41   \nl
5569.63  & Fe I   &   142  &  3.42   \nl
5662.52  & Fe I   &   77  &  4.18   \nl
5857.45  & Ca I   &   136  &  2.93   \nl
5862.36  & Fe I   &   38  &  4.55   \nl
6162.18  & Ca I   &   440  &  1.90   \nl
6166.44  & Ca I   &   45  &  2.52   \nl
6169.04  & Ca I   &   78  &  2.52   \nl
6169.56  & Ca I   &   212  &  2.52   \nl
6173.34  & Fe I   &   40  &  2.22   \nl
6261.11  & Ti I   &   38  &  1.43   \nl
6270.23  & Fe I   &   18  &  2.86   \nl
6393.61  & Fe I   &   123  &  2.43   \nl
\cutinhead{ HD140283}
4434.96  & Ca I   &   45  &  1.89   \nl
4461.66  & Fe I   &   43  &  0.09   \nl
4466.56  & Fe I   &   30  &  2.83   \nl
4468.50  & Ti II   &   77  &  1.13   \nl
4494.57  & Fe I   &   30  &  2.20   \nl
4555.89  & Fe II   &   23  &  2.83   \nl
4563.77  & Ti II   &   50  &  1.22   \nl
4871.33  & Fe I   &   36  &  2.86   \nl
4957.31  & Fe I   &   37  &  2.85   \nl
4957.61  & Fe I   &   110  &  2.81   \nl
5006.12  & Fe I   &   52  &  2.83   \nl
5012.08  & Fe I   &   31  &  0.86   \nl
5171.61  & Fe I   &   40  &  1.48   \nl
5226.87  & Fe I   &   41  &  3.04   \nl
5227.19  & Fe I   &   104  &  1.56   \nl
5232.95  & Fe I   &   46  &  2.94   \nl
5328.05  & Fe I   &   124  &  0.91   \nl
5397.14  & Ti I   &   62  &  1.88   \nl
5405.76  & Fe I   &   69  &  0.99   \nl
5424.08  & Fe I   &   26  &  4.32   \nl
5429.71  & Fe I   &   72  &  0.96   \nl
5434.53  & Fe I   &   54  &  1.01   \nl
5473.92  & Ni I   &   33  &  1.83   \nl
6162.18  & Ca I   &   36  &  1.90   \nl
\cutinhead{ $\theta$ UMa}
6213.44  & Fe I   &   62  &  2.22   \nl
6219.29  & Fe I   &   74  &  2.20   \nl
6230.74  & Fe I   &   123  &  2.56   \nl
6335.34  & Fe I   &   76  &  2.20   \nl
6336.83  & Fe I   &   80  &  3.69   \nl
6449.82  & Ca I   &   117  &  2.52   \nl
6569.22  & Fe I   &   47  &  4.73   \nl
6707.98  & Li I   &   120  &  0.00   \nl
6717.69  & Ca I   &   94  &  2.71   \nl
6843.66  & Fe I   &   35  &  4.55   \nl
7122.21  & Ni I   &   83  &  3.54   \nl
7286.52  & Ni I   &   145  &  3.77   \nl
7771.96  & O I   &   133  &  5.37   \nl
7774.18  & O I   &   122  &  4.47   \nl
7775.40  & O I   &   97  &  4.47   \nl
7780.56  & Fe I   &   84  &  4.47   \nl
\enddata
\end{deluxetable}

\begin{deluxetable}{rrrrrrr}
\scriptsize
\tablecolumns{6}
\tablewidth{0pc}
\tablecaption{Flux Mean Bisectors: Gmb1830 and the Sun (skylight)\label{table3}}
\tablehead{
\colhead{} &  \multicolumn{3}{c}{Gmb1830} &    
\multicolumn{3}{c}{Sun} \\
\cline{2-4} \cline{5-7} \\
\colhead{Normalized Flux}    &  \colhead{Mean Velocity} & 
  \colhead{Std. Dev.}   & \colhead{Number of lines} &
    \colhead{Mean Velocity} & 
  \colhead{Std. Dev.}   & \colhead{Number of lines} \\
 \colhead{}    &  \colhead{(m s$^{-1}$)} & 
  \colhead{(m s$^{-1}$)}   & \colhead{} &
    \colhead{(m s$^{-1}$)} & 
  \colhead{(m s$^{-1}$)}   & \colhead{}  } 
\startdata
0.20  & -1.64  & 16.72  & 2     & \nodata  & \nodata  & \nodata    \nl
0.25  & -1.57  & 1.08  & 3     & \nodata  & \nodata  & \nodata    \nl
0.30  & 21.76  & 15.98  & 3      & 23.19  & 2.85  & 2    \nl
0.35  & -1.01  & 8.55  & 9   & 2.74  & 4.94  & 4    \nl
0.40  & -2.64  & 6.22  & 14   & -2.37  & 3.54  & 3    \nl
0.45  & -2.56  & 7.79  & 10   & -2.69  & 12.74  & 8    \nl
0.50  & -5.67  & 7.62  & 10   & -7.68  & 19.16  & 9    \nl
0.55  & -9.40  & 10.72  & 11   & -15.60  & 24.41  & 11    \nl
0.60  & -10.29  & 10.65  & 14   & -25.85  & 11.77  & 10    \nl
0.65  & -13.26  & 17.07  & 18   & -32.43  & 19.05  & 12    \nl
0.70  & -17.02  & 17.77  & 18    & -28.70  & 22.47  & 13    \nl
0.75  & -13.55  & 28.02  & 17   & -27.80  & 27.00  & 13    \nl
0.80  & -15.08  & 20.10  & 17   & -20.75  & 34.82  & 13    \nl
0.85  & -7.24  & 27.15  & 16   & -5.80  & 48.70  & 13    \nl
0.90  & -6.95  & 25.15  & 13   & 14.43  & 52.70  & 12    \nl
0.92  & 1.13  & 13.01  & 8   & 37.08 & 55.73  & 12    \nl
0.95  & 12.31  & 18.14  & 6   & 73.80  & 74.58  & 3    \nl
\enddata
\end{deluxetable}

\begin{deluxetable}{rrrrrrr}
\scriptsize
\tablecolumns{6}
\tablewidth{0pc}
\tablecaption{Flux Mean Bisectors: HD140283 and $\theta$ UMa \label{table4}}
\tablehead{
\colhead{} &  \multicolumn{3}{c}{HD140283} &    
\multicolumn{3}{c}{$\theta$ UMa} \\
\cline{2-4} \cline{5-7} \\
\colhead{Normalized Flux}    &  \colhead{Mean Velocity} & 
  \colhead{Std. Dev.}   & \colhead{Number of lines} &
    \colhead{Mean Velocity} & 
  \colhead{Std. Dev.}   & \colhead{Number of lines} \\
 \colhead{}    &  \colhead{(m s$^{-1}$)} & 
  \colhead{(m s$^{-1}$)}   & \colhead{} &
    \colhead{(m s$^{-1}$)} & 
  \colhead{(m s$^{-1}$)}   & \colhead{}  } 
\startdata
0.40   & \nodata & \nodata & \nodata & 6.41  & 0  & 1      \nl
0.45  & 15.87 & 9.42 &  2 & -3.99  & 0  & 1     \nl
0.50  & 17.13 & 19.78 &  4 & -9.84  & 0  & 1       \nl
0.55  & -5.55 & 6.76 & 6 & 0.64  & 0  & 1       \nl
0.60  & 1.28 & 11.76 &  7 & 2.08  &  0  &  1       \nl
0.65 & 9.08 & 15.83  &  8 & -16.39 &  37.50 & 2    \nl
0.70  & 20.85 & 16.79 & 12 &   6.54  & 8.30 &  3  \nl
0.75  & 17.81  & 23.98 & 16 & 1.13 &  24.45 &  8   \nl
0.80  & 37.22 & 37.29 &  18 & 3.54 &  18.51 &  10    \nl
0.85 & 74.98 & 38.16 & 19 & 22.01 &  30.14 &  13   \nl
0.90  & 157.81 & 41.93 & 17 & 34.71 &  60.45 &  14    \nl
0.92 & 200.78 & 65.57  & 23 & 63.91 & 75.57 &  13    \nl
0.95  & 294.44 &  77.72 &  20 & 100.35 &  90.27 &  10    \nl
\enddata
\end{deluxetable}

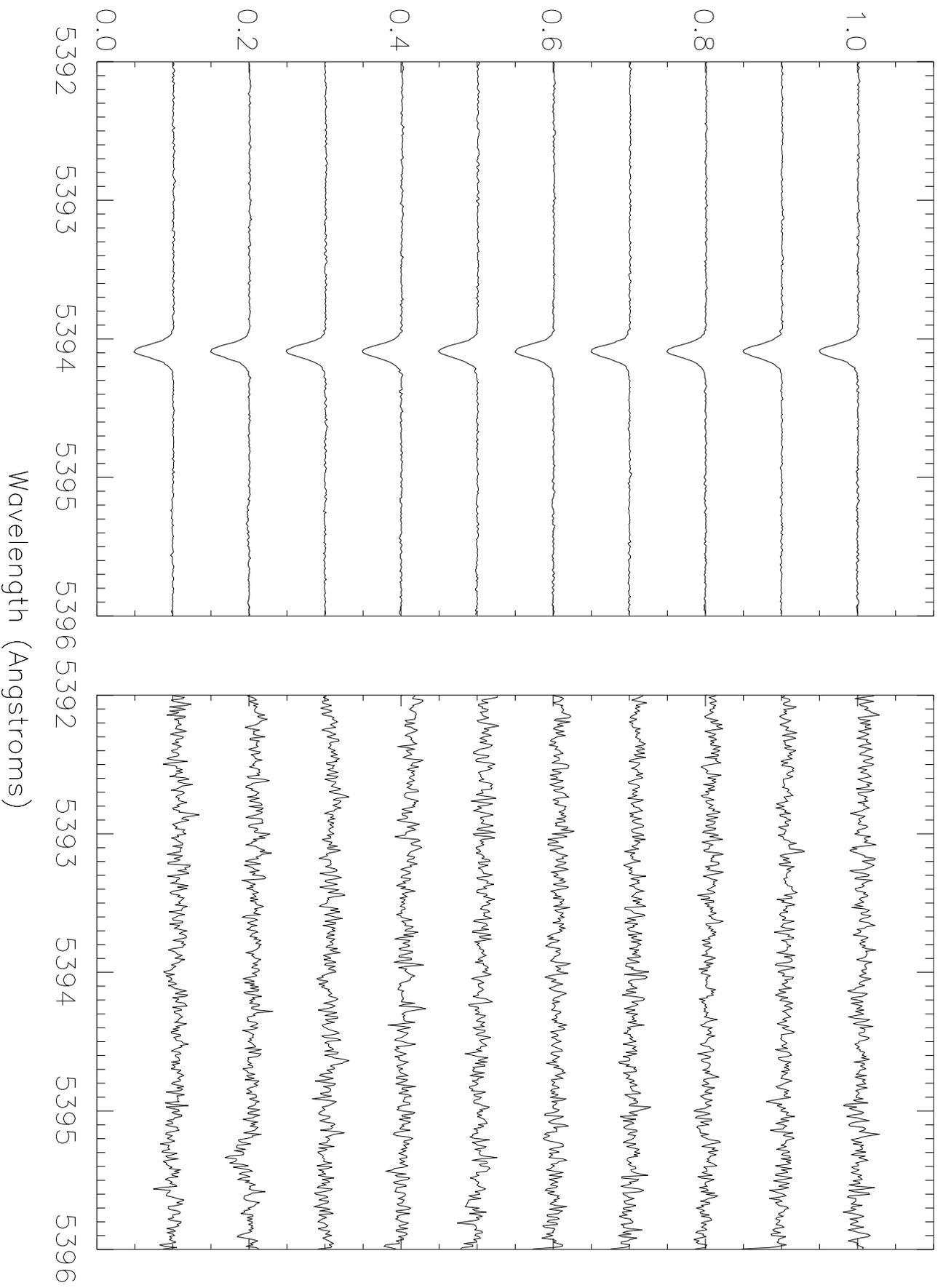
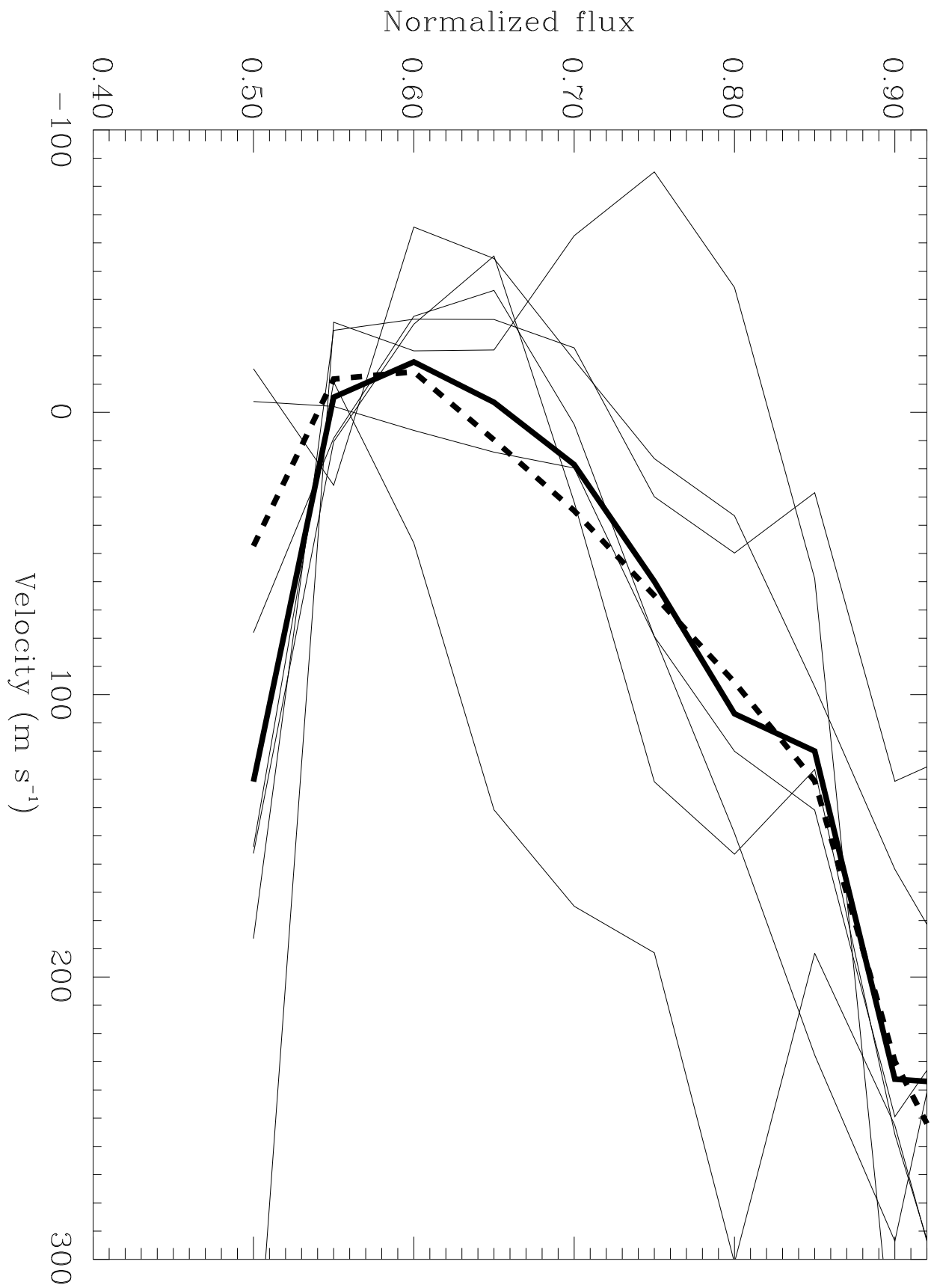
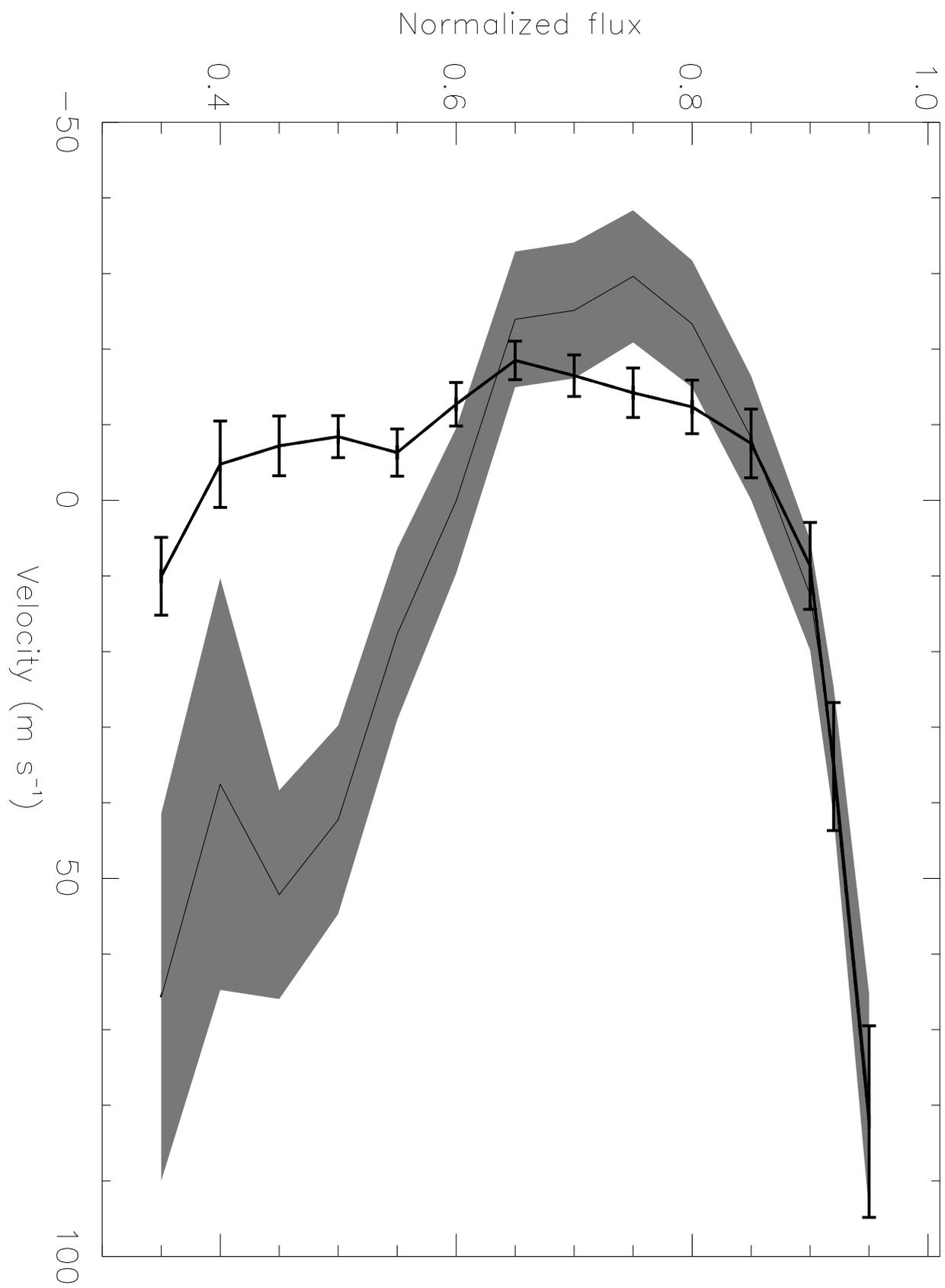
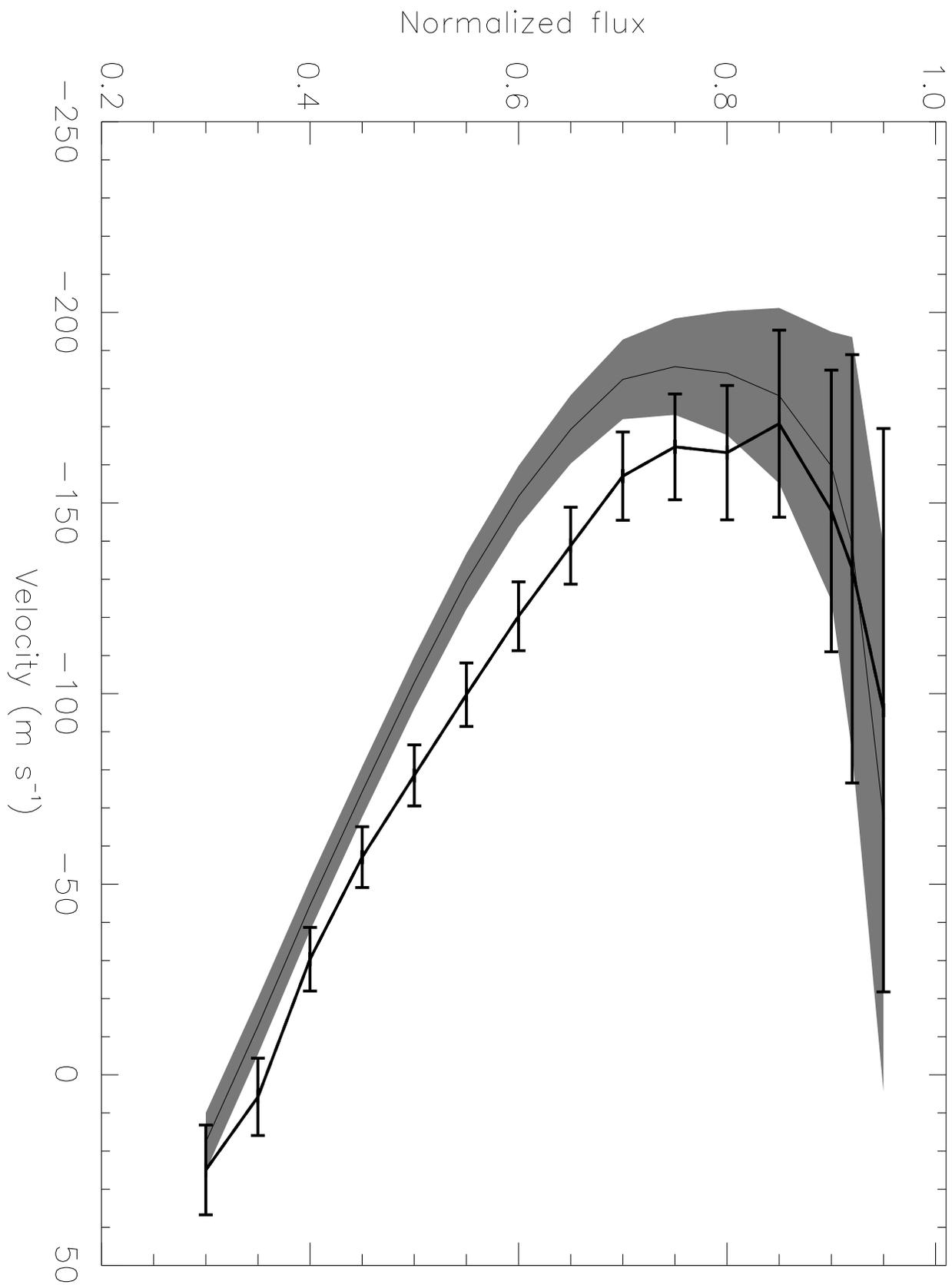
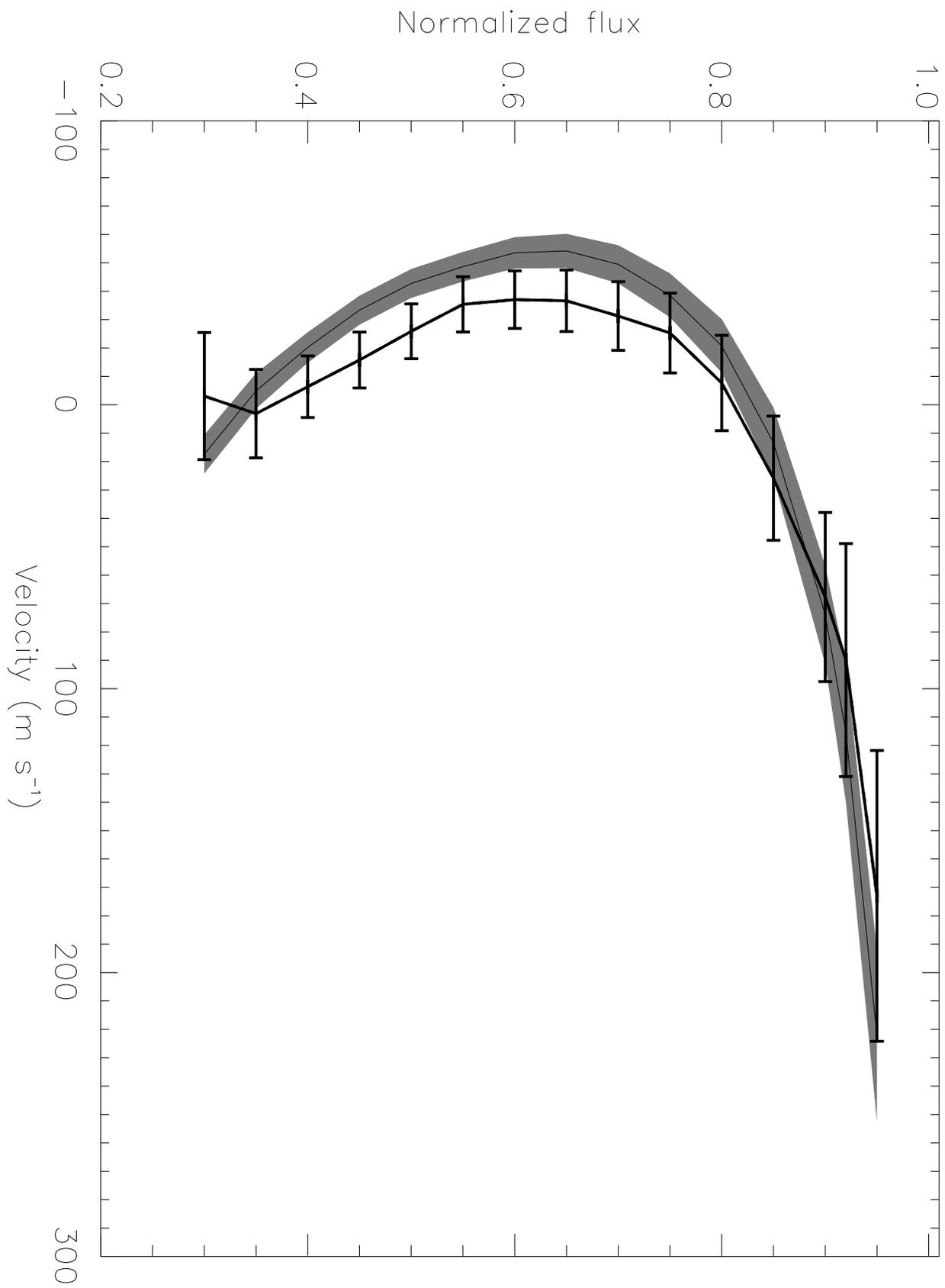
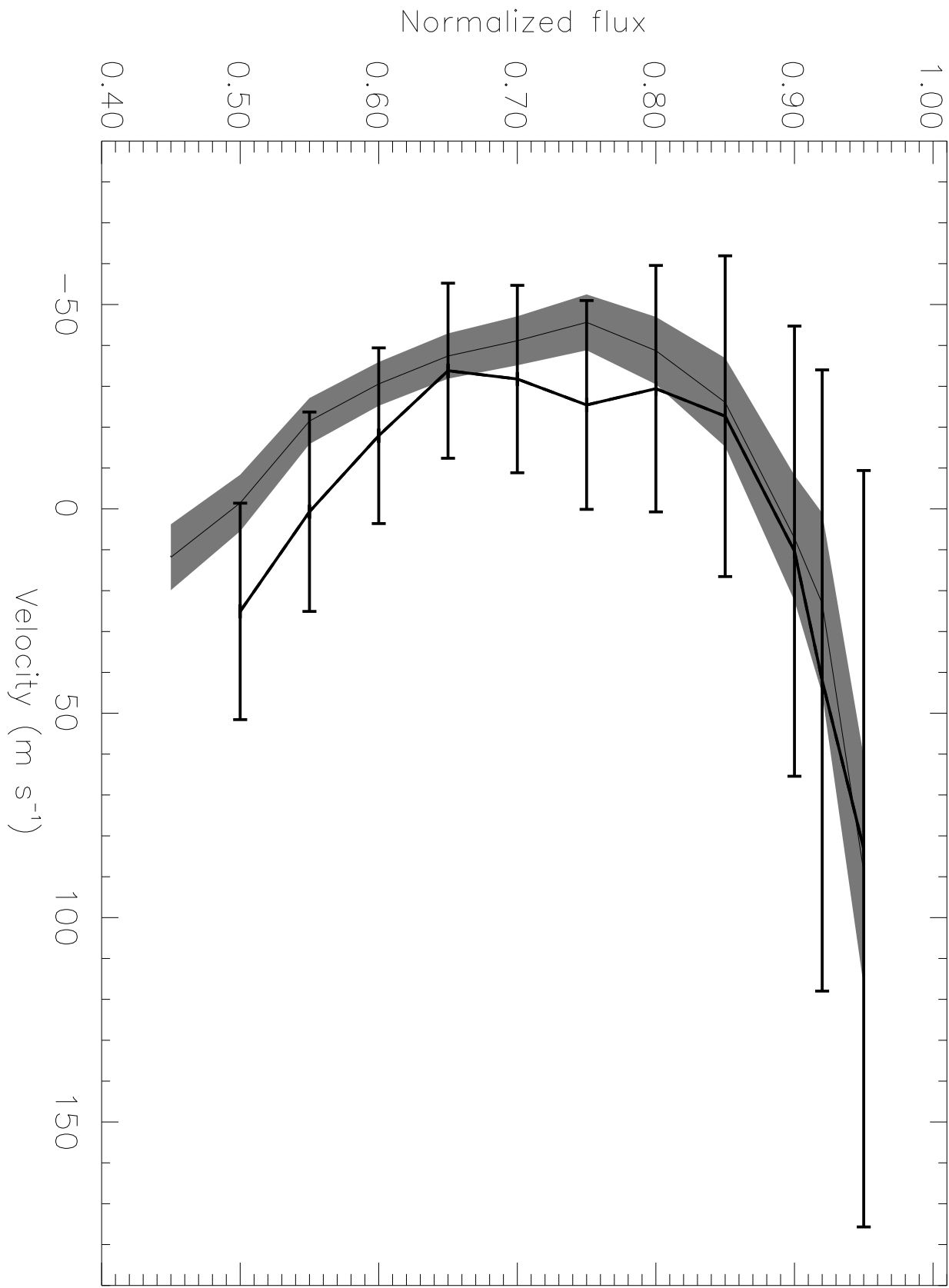
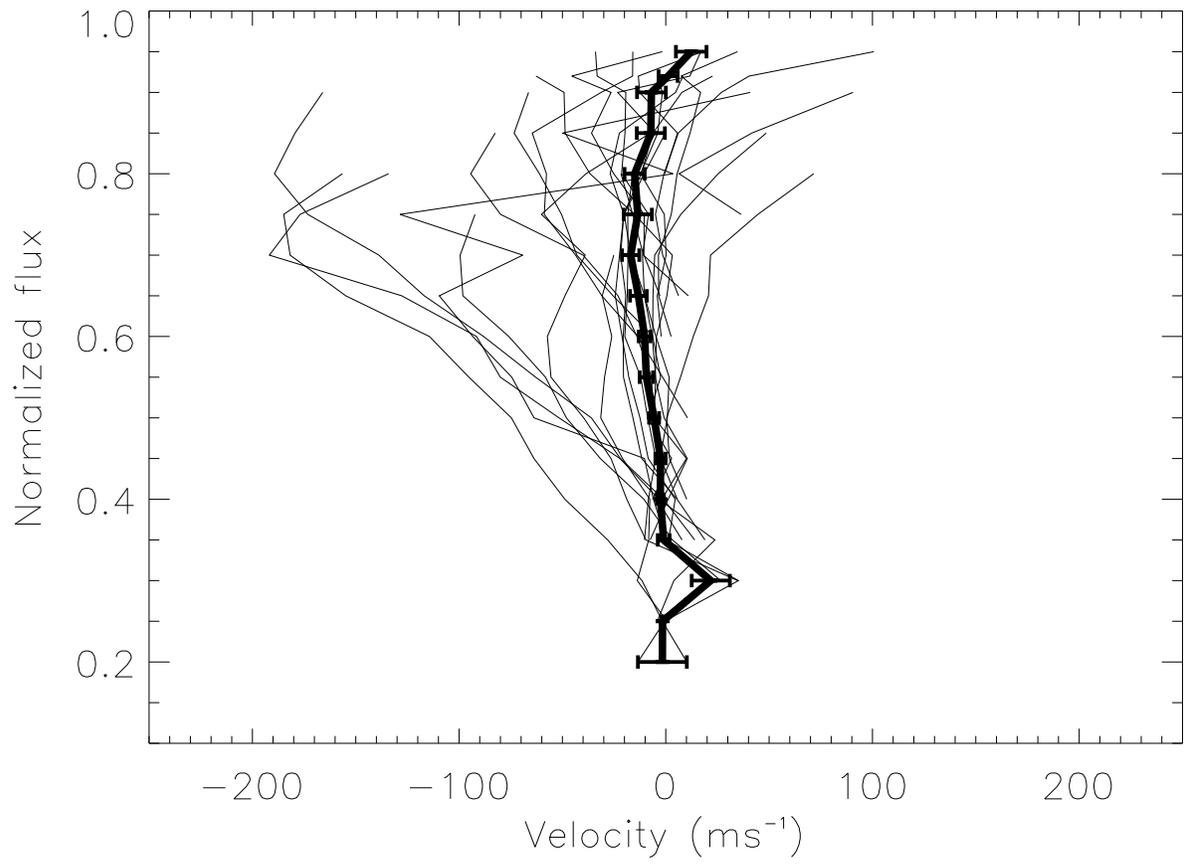
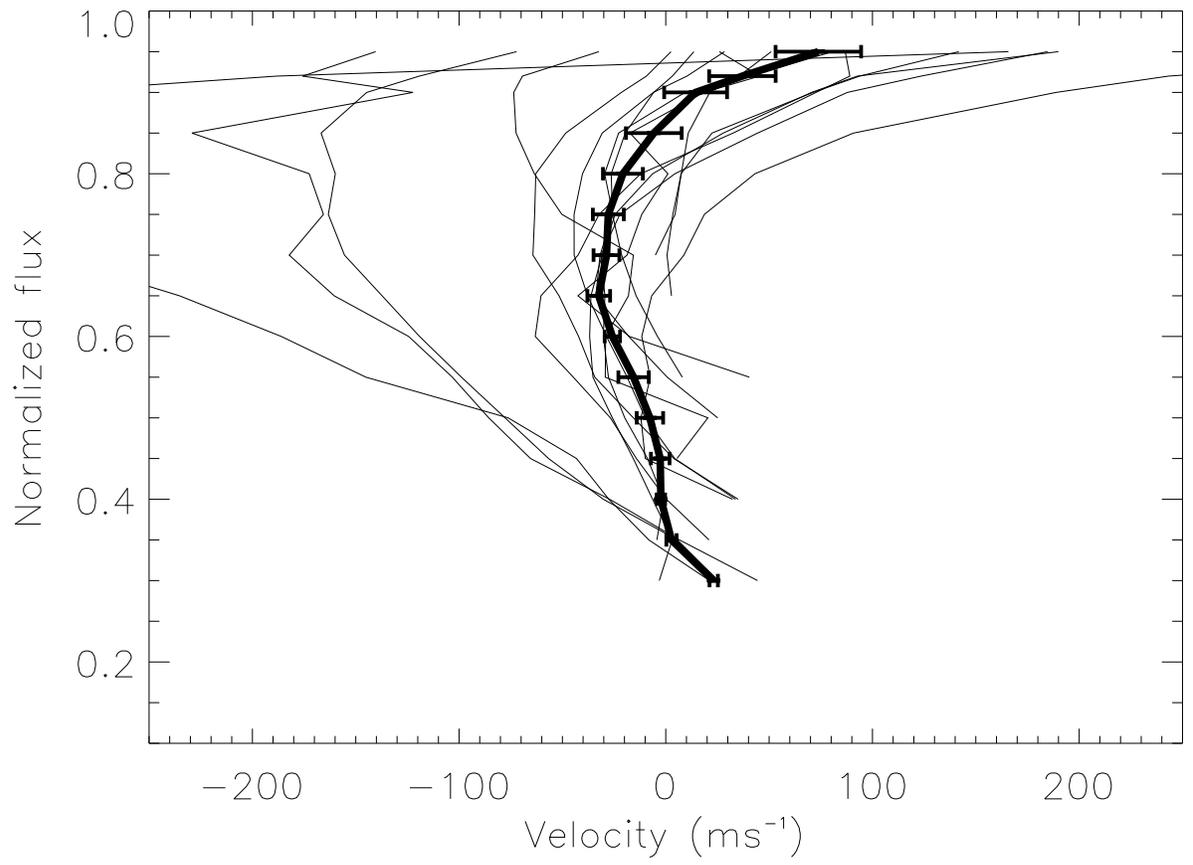
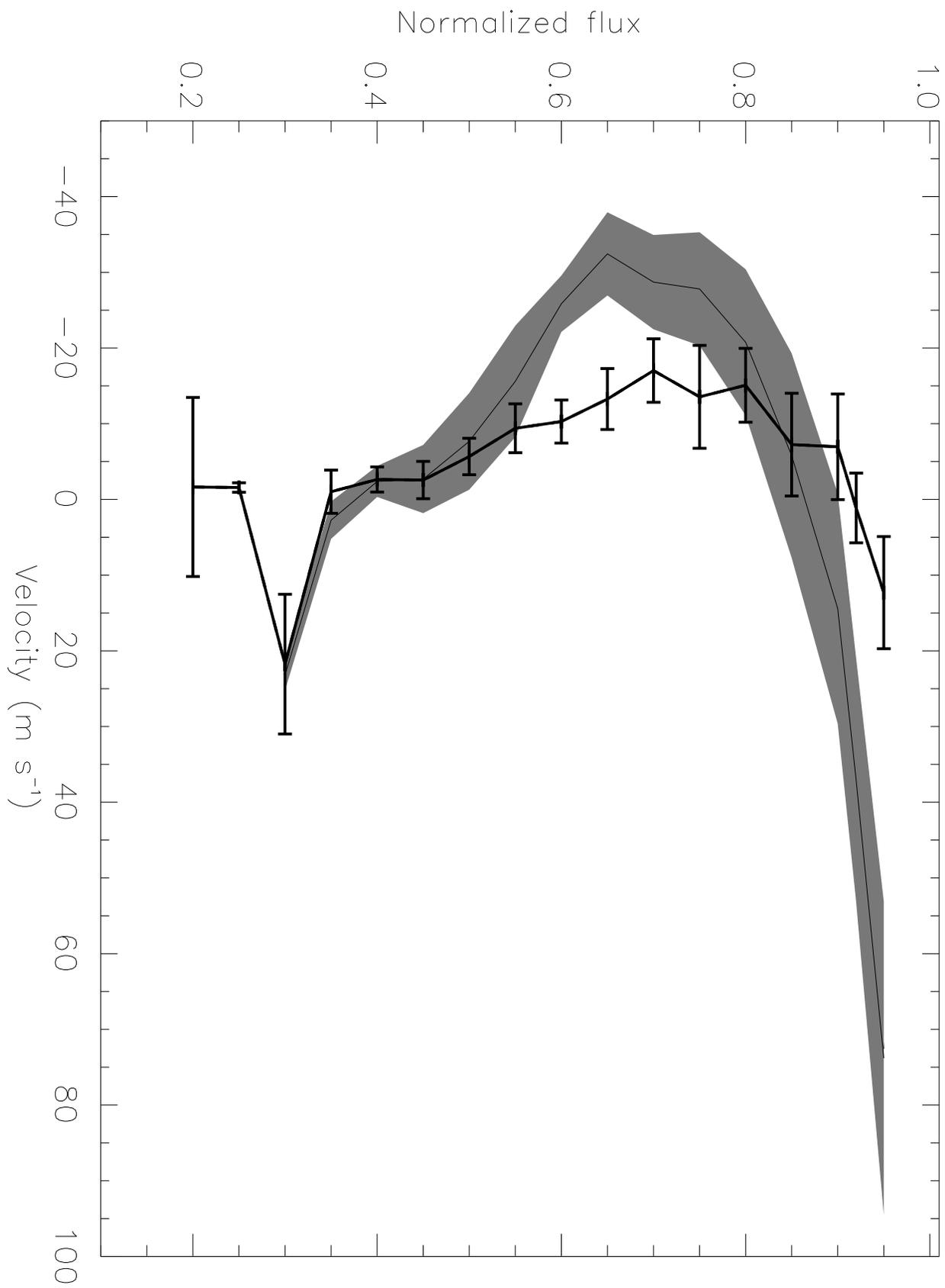
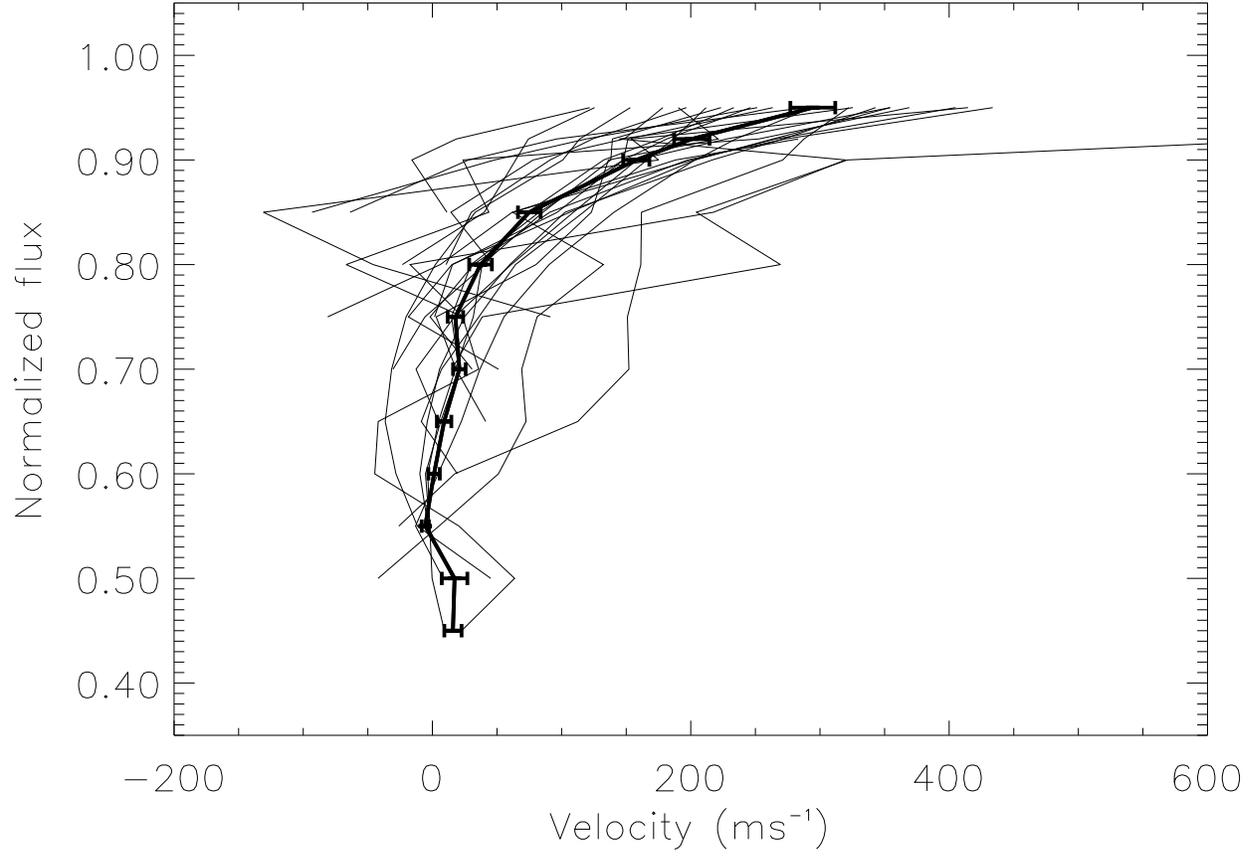
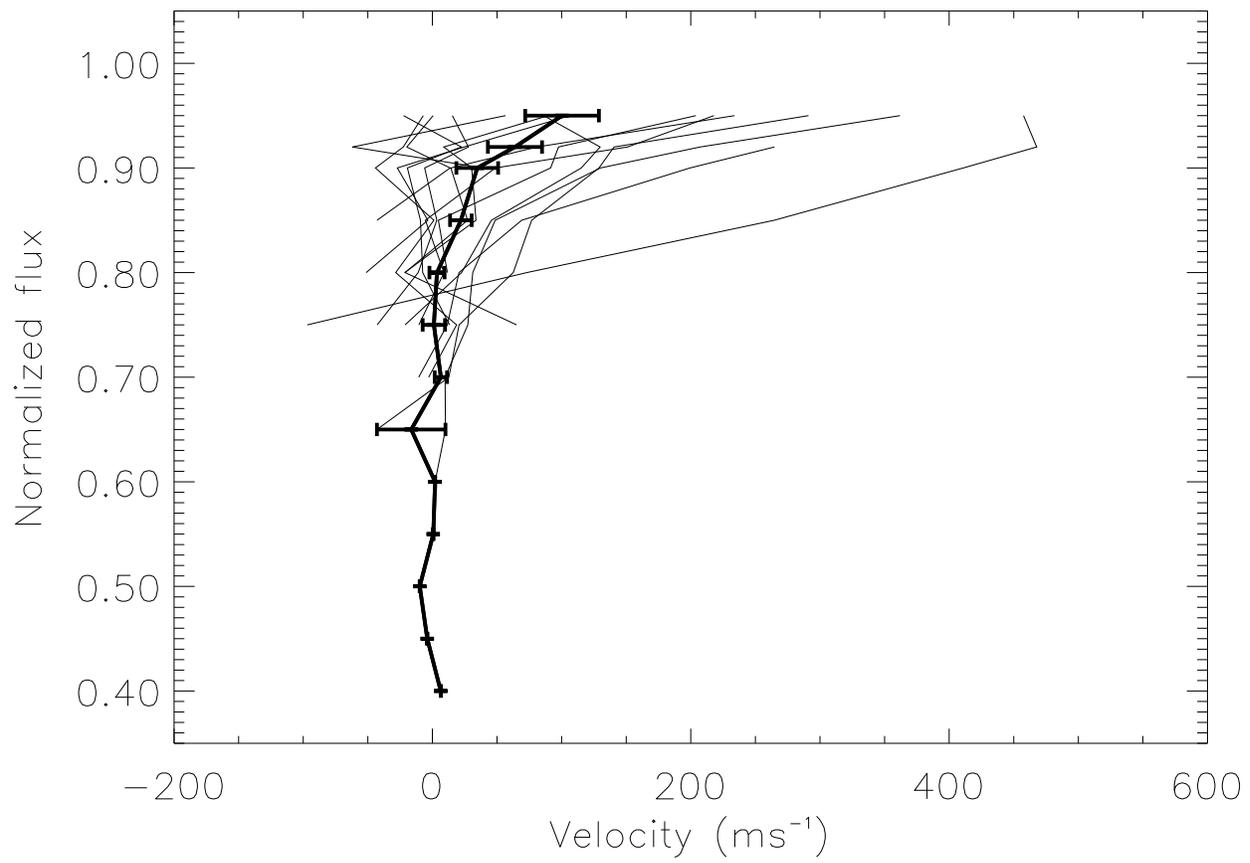
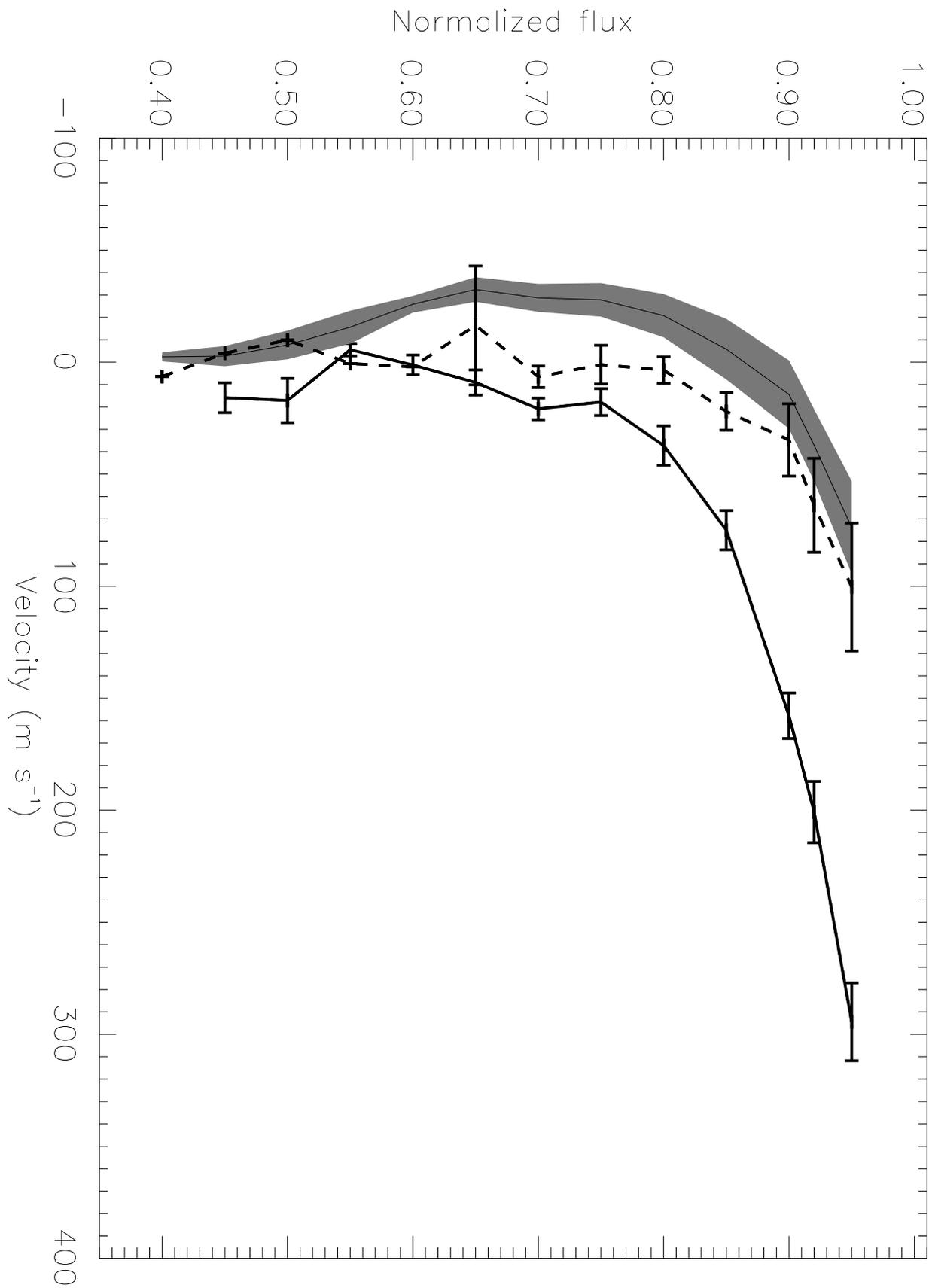
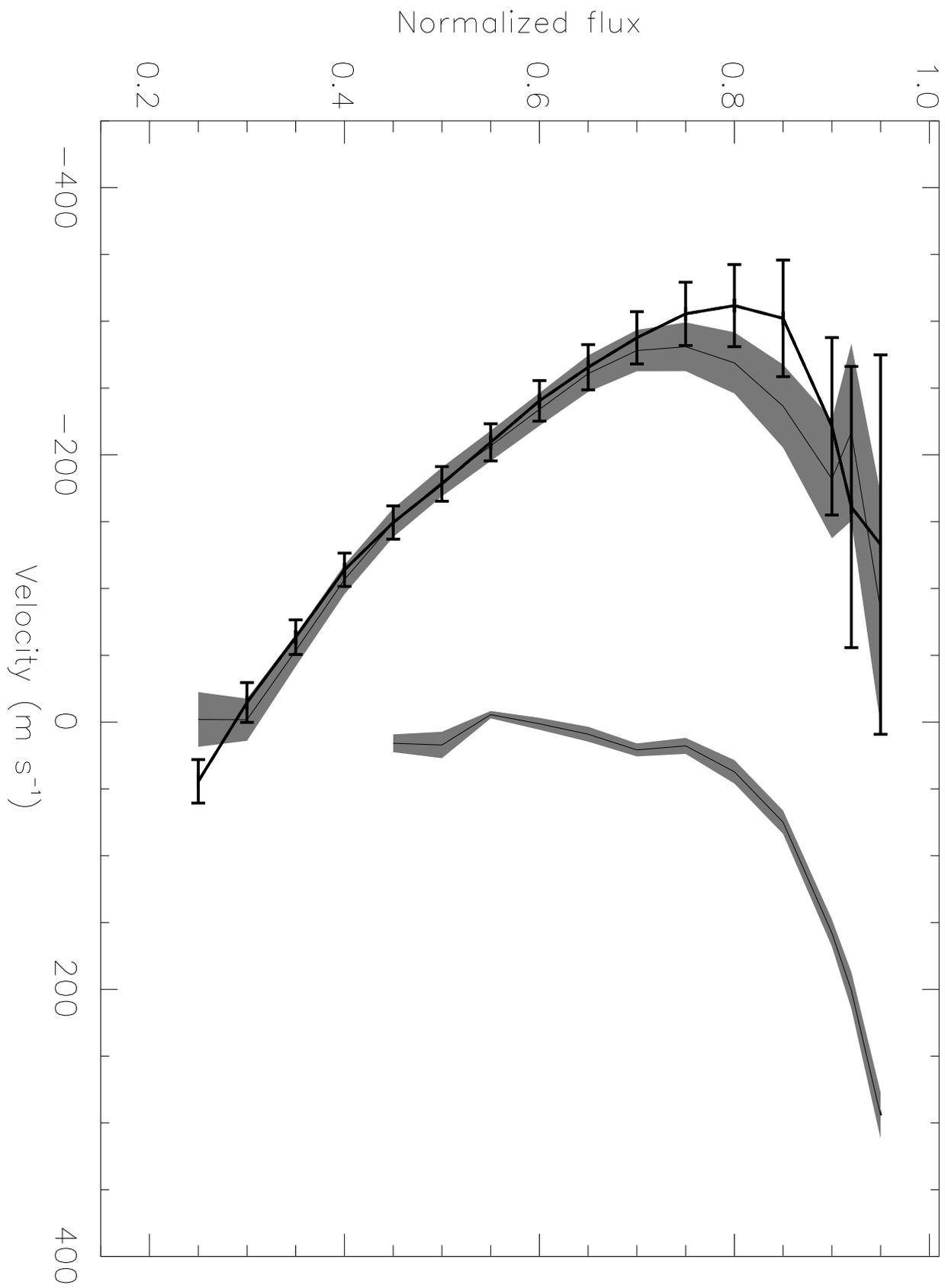
\end{document}